\documentclass[twocolumn]{jpsj3}

\title{Single Crystal Growth and Characterization of the Iron-Based Superconductor KFe$_{2}$As$_{2}$ Synthesized by KAs Flux Method}

\author{
Kunihiro Kihou$^{1, 4}$\thanks{E-mail address: k.kihou@aist.go.jp} , Taku Saito$^{2}$, Shigeyuki Ishida$^{3,4}$, Masamichi Nakajima$^{3,4}$, Yasuhide Tomioka$^{1,4}$, Hideto Fukazawa$^{2,4}$, Yoh Kohori$^{2,4}$, Toshimitsu Ito$^{1, 4}$, Shin-ichi Uchida$^{3,4}$, Akira Iyo$^{1, 4}$, Chul-Ho Lee$^{1, 4}$, Hiroshi Eisaki$^{1, 4}$}
\inst{
$^{1}$National Institute of Advanced Industrial Science and Technology(AIST), Tsukuba, Ibaraki 305-8568, Japan\\
$^{2}$Department of Physics, Chiba University, Chiba 263-8522, Japan\\
$^{3}$Department of Physics, Tokyo University, Tokyo 113-0033, Japan\\
$^{4}$JST, Transformative Research-Project on Iron Pnictides (TRIP), Tokyo 102-0075, Japan \\
}
\abst{Centimeter sized platelet single crystals of KFe$_{2}$As$_{2}$ were grown using a self-flux method. An encapsulation technique using commercial stainless steel container allowed the stable crystal growth lasting for more than 2 weeks. Ternary K-Fe-As systems with various starting compositions were examined to determine the optimal growth conditions. Employment of KAs flux led to the growth of large single crystals with the typical size of as large as 15 mm $\times$ 10 mm $\times$ 0.4 mm. The grown crystals exhibit sharp superconducting transition at 3.4 K with the transition width 0.2 K, as well as the very large residual resistivity ratio exceeding 450, evidencing the good sample quality. }

\kword{Fe-based superconductor, KFe$_{2}$As$_{2}$, single crystal growth \ldots}

\begin{document}
\maketitle

\section{Introduction}

Soon after the discovery of the superconductivity at 26 K in LaFeAs(O,F)\cite{kamihara}, studies on the iron (Fe) -based superconductors have become one of the main research subjects in the field of material science. To date, variety of superconductors, such as $Ln$FeAs(O,F) ($Ln$ : rare earth),  (Ba,K)Fe$_{2}$As$_{2}$, Fe(Se,Te), LiFeAs, $etc$., have been found and the superconducting critical temperature ($T_c$) exceeds 50 K, marking the highest record besides cuprates. Worldwide investigations towards developing novel high-$T_c$ superconductors, as well as towards elucidating their superconducting mechanism, are now extensively in progress. \cite{ishida}

Among various Fe-based superconductors, KFe$_{2}$As$_{2}$ stands out as a unique material. It is the end member of the Ba$_{1-x}$K$_{x}$Fe$_{2}$As$_{2}$ solid solution system which exhibits the highest $T_c$ of 38 K at $x$ = 0.4\cite{rotter,chen,sasmal}. Therefore one can consider KFe$_{2}$As$_{2}$ as a 'parent' compound like LaFeAsO or BaFe$_{2}$As$_{2}$, in that high-$T_c$ superconductivity emerges upon its chemical substitution. However, while other parent compounds are all non-superconducting and exhibit long-range antiferromagnetic order, KFe$_{2}$As$_{2}$ is a superconductor in itself, with relatively low $T_c$ of 3.8 K. The formal valence of Fe ions is 2.5+, much higher than those of other Fe-based superconductors which are more or less 2+. As a result, the shape of the Fermi surface (FS) is expected to be significantly modified. Band structure calculations predict that while most of the Fe-based superconductors possess two-dimensional electron- and hole-FS sheets with roughly equal size, FS's of KFe$_{2}$As$_{2}$ are dominated by large hole sheets centered at  $\Gamma$ point\cite{singh}. This feature is experimentally confirmed by the angle-resolved photoemission spectroscopy (ARPES)\cite{sato} and the de Hass-van Alphen (dHvA)\cite{terashima} experiments. Different FS shape would lead the pairing interaction distinct from other Fe-based superconductors. Indeed, recent NMR\cite{fukazawa}, penetration depth\cite{hashimoto}, and thermal conductivity\cite{dong} measurements suggest that the superconducting gaps in KFe$_{2}$As$_{2}$ have nodes, in contrast to the nodeless gaps termed s$\pm$ wave proposed for the typical Fe-based superconductors\cite{hashimoto2,hashimoto3,malone,ding,h.kim,ni}, such as $Ln$FeAsO$_{1-x}$F$_{x}$, (Ba,K)Fe$_{2}$As$_{2}$, and Ba(Fe,Co)$_{2}$As$_{2}$, $etc$. It is also intriguing that the electronic specific heat is quite large\cite{fukazawa}, suggesting the strong electron correlation and heavy carrier masses, which is indeed confirmed by the dHvA measurements\cite{terashima}. 

In order to sort out the intrinsic properties of materials, systematic investigations using sizable, high-quality single crystals are indispensable. In the case of BaFe$_{2}$As$_{2}$-related compounds, flux method is commonly used to grow single crystals. There are numbers of reports on the crystal growth of (Ba,K)Fe$_{2}$As$_{2}$ and Ba(Fe,$M$)$_{2}$As$_{2}$ ($M$ = Co, Ni, Rh, Pd, Ru) by employing tin (Sn)\cite{ni,j.s.kim,reuvekamp} flux or FeAs\cite{luo,ni2,wang,ni3,morinaga,sefat} (self-) flux. As for KFe$_{2}$As$_{2}$, several groups have grown single crystals using FeAs flux.\cite{dong,terashima2} The samples show superconducting transition at 3.0 K, slightly lower than $T_c$ of 3.8 K reported on polycrystalline samples. The in-plane residual resistivity ratio (RRR) of these crystals, defined by $\rho_{ab}$($\sim$300 K)/$\rho_{ab}$(4 K), is estimated to be 86-87.

It is known that the single crystals grown using Sn flux contain several \% of Sn contamination, which drastically affects their physical properties. For example, in the case of BaFe$_{2}$As$_{2}$, Sn contamination results in the decrease of its structural/antiferromagnetic phase transition temperature from 138 K to 80 K.\cite{ni} Accordingly, self-flux method is more favorable in growing single crystals with better quality.  In the particular case of KFe$_{2}$As$_{2}$, crystal growth using FeAs self-flux is also expected to be difficult because of hard controlling of potassium (K) at high temperature.  The vapor pressure of K becomes high around the melting temperature of FeAs (T$_{m}$ = 1030 $^{\circ}$C), and in some cases attacks the sample container.  As a result, stable single crystal growth cannot be sustained, which should degrade the size and the properties of the grown crystals.
 
A possible solution to avoid such difficulty is to employ adequate self-flux which allows the crystal growth at lower temperatures. Here in this study we have tested two kinds of self-flux with low T$_{m}$, namely, K (T$_{m}$ = 63.7 $^{\circ}$C) and KAs (T$_{m}$ = 625 $^{\circ}$C), expecting that their lower T$_{m}$ contributes for the stable synthesis of single crystal samples.  Also, in order to prevent the corrosion between the sample container and the K vapor, we have developed original encapsulation technique using a stainless steel sample container assembly. After some trials, we have found that the mixture of KAs and FeAs precursors with the mixing ratio KAs : FeAs = 3 $\sim$ 6 : 1 is most effective in growing large single crystals with the typical size 15 mm  $\times$ 10 mm $\times$ 0.4 mm. The crystals have sharp superconducting transition at 3.5 K, and large RRR exceeding 500, evidencing the good sample quality. 

\section{Experimental methods}

\begin{figure}
\begin{center}
\includegraphics[width=\columnwidth]{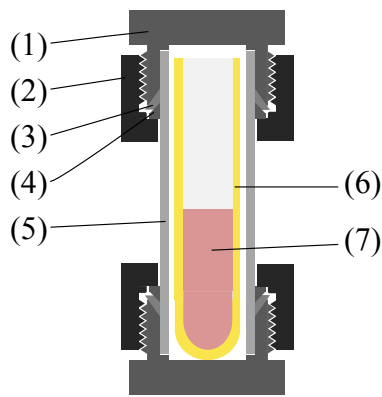}
\end{center}
\caption{(Color online) Sealing assembly for the single crystal growth. (1) cap union body (1h inner diameter)   (2) cap union nut  (3) front ring (4) back ring (5) stainless tube (1h diameter) (6)alumina crucible (7) starting materials. }
\label{f1}
\end{figure}

In this study, single crystals of KFe$_{2}$As$_{2}$ were grown using K-rich self-flux in a sealed environment. For the growth of BaFe$_{2}$As$_{2}$-related crystals, quartz glass tube is usually used to seal the samples. However, quartz glass tube is not adequate in the present case since it suffers significant chemical reaction with the evaporated K during the crystal growth. To avoid the reaction between the K vapor and the container, we have employed an encapsulation method using a commercial stainless steel container as illustrated in Fig. 1. Here an alumina crucible containing the starting materials is put into a steel tube. Tight sealing is accomplished by two caps also made of stainless steel which are screwed to the both ends of the tube. The container bears the temperature up to 1000 $^{\circ}$C. The whole assembly can be obtained commercially.

The advantages of this encapsulation method are as follows. (1) Unlike quartz tube, stainless steel does not react with K vapor. Accordingly, long-time crystal growth can be sustained. (2) Sealing process is simple and easy. In particular, welding is not required to seal the container. As a result, whole preparation processes can be carried out inside the globe box, without exposing the reactive starting materials (K/KAs) in air. (3) The assembly is relatively cheaper compared to other sealing methods. 

The starting materials are K(3N, chunk), Fe(4N, 150 $\mu$m mesh), and As(6N, 1 - 5 mm chunk). In some cases we used pre-reacted FeAs and KAs as starting materials. FeAs was synthesized by firing the mixture of Fe and As at 900 $^{\circ}$C for 10 hours in a sealed quarts tube. KAs was synthesized by firing the mixture of K and As at 650 $^{\circ}$C for 10 hours using a stainless steel container assembly. 

\begin{figure}
\begin{center}
\includegraphics[width=\columnwidth]{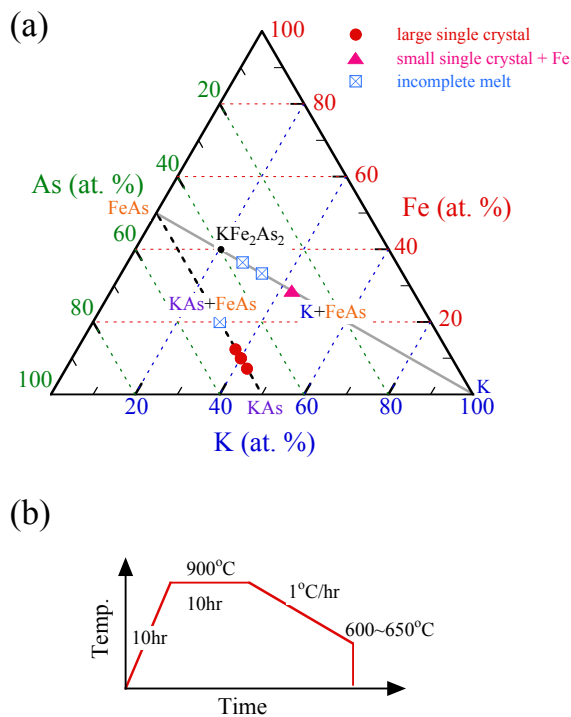}
\end{center}
\caption{(color online) (a) K-Fe-As triangular diagram for the compositions of the starting materials to grow KFe$_{2}$As$_{2}$ single crystals. The compositions surveyed in the present study are indexed. (b) Temperature profile of the single crystal growth.}
\label{f2}
\end{figure}

Typically 30g of the mixtures were put into an alumina crucible. Then the alumina crucible was sealed in a stainless steel container. The process is carried out within a globe box filled with dried N$_{2}$. The humidity and the oxygen content inside the globe box are kept below 1ppm. The sample was placed into a box furnace and fired up to 900 $^{\circ}$C for 10 hr. The temperature was kept for 10 hr, and then slowly cooled down to 600 $\sim$ 650 $^{\circ}$C at a cooling rate 1 $^{\circ}$C/hr (lasting for 10 days), as shown in Fig.2 (b). After the crystal growth, flux was rinsed out of the samples using ethanol. 

The composition of the grown single crystals was checked by the energy dispersive X-ray analysis equipped with the scanning electron microscopy (SEM; JEOL JSM-6301F, EPMA; EDAX Phoenix 1M). Crystallinity of the single crystals was evaluated by X-ray Laue photograph using a tungsten target. Lattice parameters were checked by the X-ray diffraction (XRD) patterns using CuK$\alpha$ radiation (Rigaku ULTIMA IV). Magnetic susceptibility was measured using a SQUID magnetometer (Quantum Design MPMS) under a magnetic field of 10 Oe.  The dc resistivity was measured by conventional four-probe method using Quantum Design PPMS.

\section{Results and discussion}
\subsection{Determination of the optimal growth composition}
In Fig. 2(a), we summarize the compositions we have examined to grow single crystals. The gray solid line corresponds to the solid solution of K and FeAs. Here KFe$_{2}$As$_{2}$ is located on this line at K = 20 (at.) \%. Along this line, K less than 20 \% corresponds to the FeAs self-flux growth condition and K more than 20 \% corresponds to the K self-flux growth condition, respectively. In this study we employed the compositions K : FeAs = 1.5 : 2, 2 : 2, and 3 : 2, which equal to K = 27, 33, and 43 \%, respectively. Among them, in the case of K = 27  \% and 33 \%, melting was incomplete and the resultant material was a sponge-like ingot with a lot of voids. No single crystals were recognized. For K = 43 \%, complete melting was accomplished and the resulting material is a dense ingot containing KFe$_{2}$As$_{2}$ single crystals with their typical dimension less than 1 mm $\times$ 1 mm $\times$ 0.1 mm. From the magnetic susceptibility and XRD measurements, it is found that the single crystals contain ferromagnetic Fe impurities. These impurities are mainly located at the grain boundaries and cannot be separated from the crystals. The most likely reason for the formation of Fe impurities is that the reaction,  Liq. (K) + Solid (FeAs) $ \rightarrow $ Liq. (K+As) +Solid (Fe), takes place while heating and the resultant Fe does not dissolve in the flux during the crystal growth.
 
In order to prevent the possible reaction, we have then employed the flux which contains enough amount of K+As from the beginning. The dashed line in Fig.2 (a) represents the solid solution of FeAs and KAs. In this study we examined the mixture of KAs and FeAs with the molar ratio 1.5 : 1, 3 : 1, 4 : 1, and 6 : 1, which corresponds to the K concentration of 30, 38, 40, and 43 \%, respectively. Among them, K concentration larger than 38 \% yielded the complete melting. KFe$_{2}$As$_{2}$ single crystals can be found inside the grown ingot. The size of the crystals increases with increasing the concentration of KAs flux, reaching 15 mm $\times$ 10 mm $\times$ 0.4 mm in the case of KAs : FeAs = 6 : 1. The photograph of the grown crystal is shown in Fig. 3(a) together with a millimeter scale. SQUID and XRD measurements indicate that the grown crystals are free from Fe impurities. We therefore used these crystals for the characterization and the physical property measurements.

\subsection{Crystallographic characterization}

\begin{figure}
\begin{center}
\includegraphics[width=\columnwidth]{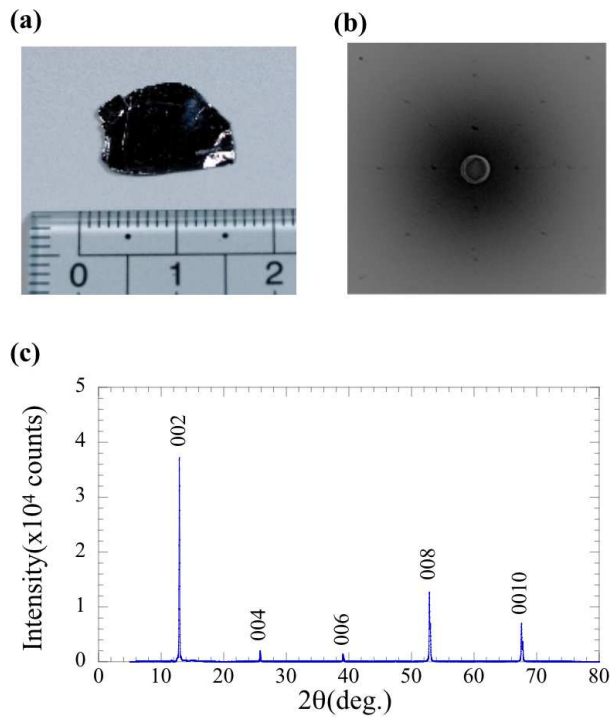}
\end{center}
\caption{(color online) (a) Photograph of the KFe$_{2}$As$_{2}$ single crystal together with a millimeter scale. (b) Back reflected Laue Photograph of KFe$_{2}$As$_{2}$. The incident x-ray beam was applied normal to the cleaved surface. (c) X-ray diffraction pattern of the KFe$_{2}$As$_{2}$ single crystal. Only (00$l$) peaks are recognized. }
\label{f3}
\end{figure}

As seen in Fig. 3(a), the grown crystals exhibit plate-like morphology with flat, shinny, black-colored surfaces. The sample easily cleaves along these surfaces. X-ray Laue backscattering photograph from the cleaved surface is shown in Fig.3(b). Four-fold symmetry is easily recognized, indicating that the flat surface corresponds to the (001) plane of the tetragonal KFe$_{2}$As$_{2}$ unit cell. Fig. 3(c) shows the x-ray diffraction pattern with the (00$l$) reflections. The $c$-axis lattice constant $c$ = 13.88 \AA\ was calculated from the observed peaks, consistent with the literature value reported on the polycrystalline samples.\cite{rotter}

According to the microprobe analysis, the composition of the grown crystal is estimated as K : Fe : As = 20.84 :  37.57 : 41.59, in accordance with the stoichiometric composition of 1 : 2 : 2 within the experimental accuracy. The result suggests that there is no appreciable K deficiency in our single crystals.

\begin{figure}
\begin{center}
\includegraphics[width=\columnwidth]{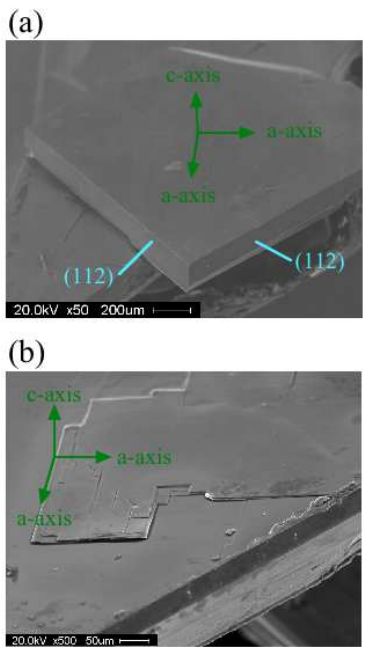}
\end{center}
\caption{SEM micrographs of KFe$_{2}$As$_{2}$. (a) Typical single crystal possessing well developed facets. (b) (001) face after the cleavage. Multilayer stacks with edges arraying along (100) direction are recognized.}
\label{f4}
\end{figure}

In Fig. 4(a) and (b), we show the SEM micrographs of the single crystal. As seen in Fig. 4(a), the crystal exhibits the tetragonal shape, reflecting the layered, tetragonal crystal structure of KFe$_{2}$As$_{2}$. The large plane corresponds to the (001) face and the edges of the crystals are identified as (112) facets, which are running along the [110] crystallization axis.  

On the (001) face, one can recognize terrace structures comprised of multiple layers with each thickness of 5ƒÊm. (Fig. 4 (b)) The terrace is formed upon the cleavage of the surface layer. The figures show that the KFe$_{2}$As$_{2}$ selectively cleaves along the (100) direction (besides the (001) direction), 45 degrees off from the growth facets running along the [110] axis.

\subsection{Physical properties}

\begin{figure}
\begin{center}
\includegraphics[width=\columnwidth]{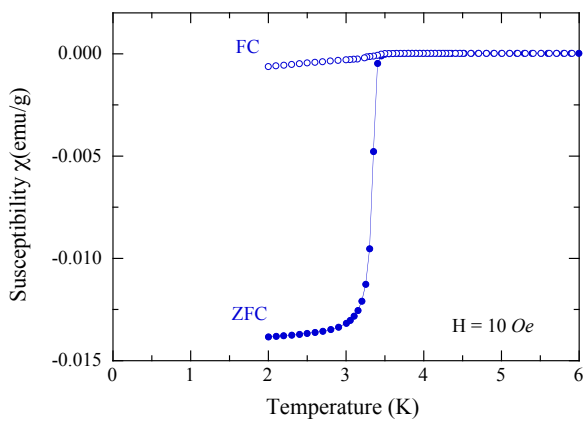}
\end{center}
\caption{Temperature dependence of the magnetic susceptibility for the KFe$_{2}$As$_{2}$ single crystal. Measurement was done under the magnetic field of 10 Oe applied perpendicular to the $c$-axis.}
\label{f5}
\end{figure}

Fig. 5 shows the temperature dependence of the zero-field-cooling (ZFC) and the field-cooling (FC) magnetic susceptibility ($\chi$) under the magnetic field of 10 Oe applied perpendicular to the $c$-axis. Both ZFC and ZF curvature show well-defined drop at 3.4 K, indicative of the onset of superconducting transition at this temperature. The 10 \%-90 \% transition width is as narrow as 0.2 K. The $T_c$ value and the superconducting transition width of the present sample is superior to those reported in literatures\cite{dong,terashima2}. The volume fraction estimated from the ZFC value below 3 K is approximately 100 \%, indicating the bulk nature of the superconductivity. Sharp superconducting transition, higher $T_c$, and large volume fraction signify the good quality of the present samples. We also note that $T_c$ is precisely reproducible. There is no appreciable difference in $T_c$ (within 0.2K) among the grown samples. The signal above $T_c$ is nearly zero, indicating that the sample is free from magnetic impurities.

\begin{figure}
\begin{center}
\includegraphics[width=\columnwidth]{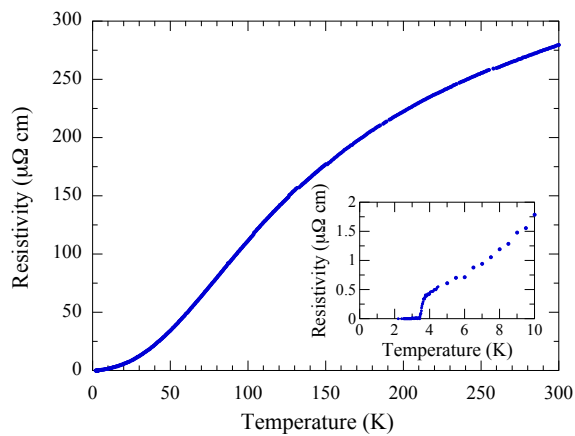}
\end{center}
\caption{Temperature dependence of the in-plane dc resistivity for the KFe$_{2}$As$_{2}$ single crystal. Inset shows the same data around the superconducting transition. }
\label{f6}
\end{figure}

Fig. 6 shows the typical in-plane dc resistivity ($\rho_{ab}$($T$)) of the grown crystal. The resistivity shows metallic ($d\rho_{ab}$/dT$>$ 0) behavior in the entire temperature range. The residual resistivity ratio (RRR), which is defined by the  $\rho_{ab}$(300 K) / $\rho_{ab}$(5 K), is as large as 458, indicating that the present sample is extremely clean, almost free from elastic impurity scattering. The number is much larger than the value of 87-88 reported on the crystals grown by the FeAs flux method.\cite{dong,terashima2}. The low-temperature ($T<$ 30K) resistivity follows the Fermi-liquid dependence, namely,$\rho_{ab}$(T)=$\rho_{0}$+$AT^2$. Its physical implication is depicted in Ref. \cite{fukazawa} and in Ref. \cite{hashimoto}.

The electrical resistivity has decreased gradually at around 3.8 K. $T_c$(midpoint) = 3.5 K and $T_c$(zero-resistivity) = 3.4 K, respectively.  $T_c$ determined by the zero-resistivity is consistent with the onset of diamagnetic signal. Accordingly, we employ the $T_c$=3.4K defined by the magnetic susceptibility as the bulk $T_c$ of the obtained single crystals.


\section{Conclusion}
We have successfully grown centimeter-sized platelet single crystals of KFe$_{2}$As$_{2}$ using a self-flux method. An encapsulation technique using stainless steel container was developed, which allowed the stable crystal growth lasting for more than 2 weeks. Ternary K-Fe-As systems with various starting compositions have been examined to establish the optimal growth conditions. Employment of KAs flux led to the growth of large single crystals with the typical size of as large as 15 mm $\times$ 10 mm $\times$ 0.4 mm. The grown crystals exhibit sharp superconducting transition at 3.4 K with the transition width 0.2 K as well as the residual resistivity ratio exceeding 450, evidencing the good sample quality.
filamentary
\section*{Acknowledgement}
The authors thank P. M. Shirage, H. Kito, Y. Yoshida, M. Ishikado, T. Terashima, M. Kimata, S. Uji, T. Yoshida, A. Fujimori, K. Hashimoto, T. Shibauchi, Y. Matsuda, and H. Kawano-Furukawa for valuable discussions and K. Shimada for SEM measurements. This work was supported by Grants-in-Aid for Specially promoted Research (20001004) and for Scientific Research on Innovative Areas "Heavy Electrons" (No. 20102005) from the Ministry of Education, Culture, Sports, Science and Technology of Japan and by a Grant-in-Aid for Scientific Research C (No. 22540380) from the Japan Society for the Promotion of Science and by Mitsubishi Foundation.


\end{document}